\documentclass[sigconf,authorversion,nonacm,screen]{acmart}

\AtBeginDocument{%
  \providecommand\BibTeX{{%
    \normalfont B\kern-0.5em{\scshape i\kern-0.25em b}\kern-0.8em\TeX}}}

\usepackage{soul}

\copyrightyear{2024}
\acmYear{2024}
\setcopyright{rightsretained}
\acmConference[CHI EA '24]{Extended Abstracts of the CHI Conference on Human Factors in Computing Systems}{May 11--16, 2024}{Honolulu, HI, USA}
\acmBooktitle{Extended Abstracts of the CHI Conference on Human Factors in Computing Systems (CHI EA '24), May 11--16, 2024, Honolulu, HI, USA}
\acmDOI{10.1145/3613905.3650997}
\acmISBN{979-8-4007-0331-7/24/05}

\begin{document}

\title[Theorizing Deception]{Theorizing Deception: A Scoping Review of Theory in Research on Dark Patterns and Deceptive Design}



\author{Weichen Joe Chang}
 \email{weich0521@gmail.com}
\orcid{0009-0003-8140-0021}
\affiliation{%
  \institution{Tokyo Institute of Technology}
  \city{Tokyo}
  \country{Japan}
}

\author{Katie Seaborn}
\email{seaborn.k.aa@m.titech.ac.jp}
\orcid{0000-0002-7812-9096}
\affiliation{%
  \institution{Tokyo Institute of Technology}
  \city{Tokyo}
  \country{Japan}
}

\author{Andrew A. Adams}
\email{aaa@meiji.ac.jp}
\orcid{0000-0003-0410-802X}
\affiliation{%
 \institution{Meiji University}
 \city{Tokyo}
 \country{Japan}
}

\renewcommand{\shortauthors}{Chang, Seaborn, and Adams}

\begin{abstract}
The issue of dark patterns and deceptive designs (DPs) in everyday interfaces and interactions continues to grow. DPs are manipulative and malicious elements within user interfaces that deceive users into making unintended choices. In parallel, research on DPs has significantly increased over the past two decades. As the field has matured, epistemological gaps have also become a salient and pressing concern. In this scoping review, we assessed the academic work so far---51 papers between 2014 to 2023---to identify the state of theory in DP research. We identified the key theories employed, examined how these theories have been referenced, and call for enhancing the incorporation of theory into DP research. We also propose broad theoretical foundations to establish a comprehensive and solid base for contextualizing and informing future DP research from a variety of theoretical scopes and lenses.  
\end{abstract}





\maketitle


\section{Introduction and Background}

Electronic devices, from personal computers to smartphones and tablets, have become the primary ways through which people access digital content and online services. There is also a continuing and expanding problem with manipulative, disruptive, and sometimes malicious elements of user interfaces (UIs) designed to trick users into making unintended choices, referred to as dark patterns and/or deceptive designs (DPs)~\cite{gray2023dpsysreview}. Research on DPs has received attention and prompted critical discussions on ethical design practices and human-computer interaction (HCI). Many DP taxonomies have been offered to clarify different types of DPs~\cite{oecd2022,gray2018}. 
Following this research, DP awareness research conducted by Di Geronimo et al.~\cite{DiGeroimo} showed that 55\% of participants could not recognize DPs in the context of a typical app. In 2021, Bongard-Blanchy et al.~\cite{bongard-blanchy} performed an online survey, finding that 59\% of participants were able to correctly identify five or more DPs across nine interfaces. Furthermore, the results indicated differences in DP recognition based on age and educational level. As long as DPs exist, research on the conceptual and user sides will be needed.

Even so, it remains unclear what theories of deception, or those related to it, are present or implied within the field and how these theories have been utilized in research. Hence, we conducted a scoping review on theory in the DP literature. We asked: \textbf{RQ1: What theories of, or related to, deception are present or implied within the field?} and \textbf{RQ2: How have these theories been used in research, regulatory, or design practice?} 
We contribute a comprehensive summary of the state of the art on theory in DP research. We have conducted an analysis of definitions and explored how these theories have been used (or not). We identify three important issues: (1) just under half of the papers (46.8\%) did not reference a theory; (2) nearly half of the papers (46.5\%) incorporated theories without providing sufficient details; and (3) two papers referred to theories that may not genuinely possess a theoretical nature. 
Our work offers a pathway for scholars new and old to this domain, enabling them to explore avenues for future contributions within DP scholarship. Simultaneously, this work offers a better understanding of theory this emerging area of study, as well as provocations for future scholarship opportunities.


\section{Methods}

This scoping review was guided by the PRISMA Extension for Scoping Reviews (PRISMA-ScR) protocol~\cite{PRISMA_checklist}. We describe our procedure for identifying and screening papers next (summarized in Supplementary Materials Appendix A). Our protocol was registered before data analysis on December 1\textsuperscript{st} 2023 via OSF\footnote{\url{https://osf.io/3md9g}}.

\subsection{Definitions}

\subsubsection{Dark Pattern} DPs are malicious interface designs that trick, force, or induce people into taking an action that is different from what they intend or expect~\cite{Brignull_2019}. Brignull in 2010 registered darkpatterns.org to address the growing issue of deceptive design practices. Recently, there has been a shift in language around the word ``dark pattern'' in light of the negative associations it embodies between ``darkness'' and ``badness.''\footnote{\url{https://www.acm.org/diversity-inclusion/words-matter}} ``Dark'' is less accurate than ``deceptive'' and is also commonly used to describe skin tone, and thus to avoid this racialized term, many are turning to other phrases, notably deceptive design, including Brignull himself~\cite{Brignull_2019,Brignull_2023}, whose site is now \url{https://www.deceptive.design/}.
    
\subsubsection{Theory and Relevant Terms} 







The study of DPs is a multidisciplinary initiative. Authors may use different terms when referring to theory; for example, persuasion theory has been referred to as persuasion ``techniques''
~\cite{Bösch2016,wu2022malicious,cialdini2009influence}. Theory may also be defined differently across disciplines. \citet[p. 613]{Gregor2006} notes how the domain of interest for a discipline can influence the nature of its theorizing. Theory in mathematics and music, for example, means different things, with knowledge developed, specified, and used in different ways across fields. We approached this challenge through an iterative process. In the screening phase (\autoref{sec:screening}), two of us read the abstracts and scanned the full-text, searching for terms related to theory. 
We discovered five relevant terms---general theory, principle, model, framework, and mechanism---that captured the notion of theory across all papers. We report on our results below, but start by defining the terms and their varieties here.

When it comes to ``theory,'' a widely accepted definition is by Sutton and Barry~\cite[p. 378]{notheory}, who defined \emph{general theory} as the connections among phenomena, a story about why acts, events, structure, and thought occur. A \emph{principle} can be defined as an 
rule that universally regulates the formation, operation, or change of any phenomena. Principles can appear in various modes: as logical inference in scientific laws, or 
as moral principles that we follow~\cite{McDonald2009}. A \emph{model} provides a simplified description of a phenomenon, and as such is closely related to theory, as defined above~\cite{notheory}. Indeed, the distinction between a theory and a model is not always clear~\cite{nilsen2015}. A \emph{framework} provides a structure for phenomena consisting of descriptive categories, but does not offer an explanation~\cite{nilsen2015}. 
A \emph{mechanism} provides an explanation on how a phenomenon comes about or how some significant process works~\cite{Machamer2000}. 

The importance and finer differences of theory remain unknown for DP studies, but we can learn from broader disciplines of practice. \citet{abend2008theory} argues that theory is so important in the lexicon of contemporary sociology that sociologists use ``theoretical'' and ``theorize'' frequently and with weight 
Other fields of research with empirical roots have struggled in their search for adequate theory, such as Operations Research in the 1990s. In that context, Wacker~\cite[p. 362-363]{wacker1998definition} justifies the need for theory:
\begin{quote}
There are three reasons why theory is important for researchers and
practitioners: (1) it provides a framework for analysis; (2) it provides
an efficient method for field development; and (3) it provides clear
explanations for the pragmatic world.
\end{quote}
and also defines suitable theories:
\begin{quote}
Generally, academics point to a theory as being made up of four
components: (1) definitions of terms or variables, (2) a domain where the
theory applies, (3) a set of relationships of variables, and (4)
specific predictions (factual claims).
\end{quote}
The field of DP studies is young and already pluralistic. The community should work to consolidate our conceptualization of theory for clarity and future consensus-building. 

\subsection{Eligibility Criteria}
We adopted four eligibility criteria from the protocol established by Gray et al.~\cite{gray2023dpsysreview} and added one additional criterion to identify the presence or implication of theories. In total, five eligibility criteria were applied: 1) The record had to be written in English. 2) The record had to explicitly mention ``dark patterns'' in the full text. 3) The record had to be published in a journal, conference proceeding, government technical report, or a similar archival venue. 4) The record had to include at least one empirical component. 5) The record had to contain or imply theories within the field.

\subsection{Data Collection and Screening} \label{sec:screening}
We started with a data set from the systematic review conducted by Gray et al.~\cite{gray2023dpsysreview} in 2022, covering work until September 13\textsuperscript{th}, 2022. We conducted a new search spanning the period from September 13\textsuperscript{th}, 2022, to November 30\textsuperscript{th}, 2023, adhering to the established protocol delineated by Gray et al.~\cite{gray2023dpsysreview}.
We also replicated this search protocol on Google Scholar and the ACM Digital Library, using the precise search string ``dark patterns'' (enclosed within quotation marks). This yielded 3,020 results in Google Scholar and 158 results in the ACM DL, both sorted by ``relevance.'' 
The initial 250 results from Google Scholar\footnote{We may have overlooked additional relevant literature beyond the first 250 results. However, after the first 150 results, few papers showed evidence of citations and the vast majority appeared to be irrelevant to the search. Future work could explore this broader range of results to determine if we missed any important literature.} and all 158 results from the ACM DL were downloaded and formatted in BibTeX. All BibTeX entries were then imported into Rayyan.ai\footnote{\url{https://www.rayyan.ai/}}, a tool facilitating collaborative literature reviews. 38 papers with a publication date before September 13\textsuperscript{th}, 2022 were removed. Combining all results, we obtained 449 titles and abstracts. 41 duplicate records and 39 records not written in the English language were excluded. Finally, we attempted to download the full text of the remaining 369 reports, retrieving all but eleven.

Screening the remaining papers for eligibility was conducted using the five inclusion criteria stated above. 
Reports were excluded based on the following factors: papers of an ineligible publication type (e.g., popular press article, only an abstract, workshop, student thesis, pre-print; n=87); reports unrelated to computing or DPs (e.g., medical or hard science publication; n=4); reports that did not use DPs as a primary analytic or conceptual framing (e.g., only referencing the term in an introduction or conclusion; n=130); and reports that were not empirical (e.g., no new data collected or analyzed; only focus on argumentation; systematic review; n=28). The first four criteria resulted in the exclusion of 249 reports for 109. The fifth criterion, presenting or implying theories (i.e., no theory present; n=58) left 51 papers eligible for the analysis phase.

\subsection{Data Analysis}
We employed directed content analysis~\cite{contentreview} to examine the included literature in pursuit of identifying theories of deception and how these theories were utilized. Two researchers read the abstracts and content of the papers to ensure inclusion. A single paper may have included more than one theory and more than one type of theory. We developed a preliminary round of codes generated based on an initial review of more than half of dataset (27 reports). The two researchers then confirmed the relevance and scope of the preliminary codes and used them to code all 51 reports. One researcher coded all the papers, and then the other researcher confirmed the codes across the entire dataset.

\section{Results}
We present 
the theories referenced and elaborate on how these were utilized in this body of published research. The spreadsheet containing the screened and included literature can be found at \url{https://bit.ly/theorizingdeception}.

\subsection{Theories Referenced}
Five types of theories were confirmed 
in more than half of the dataset (27 reports):7.5\% (n=19) general theory, 7.5\% (n=3) principle, 15.0\% (n=6) model, 20.0\% (n=8) mechanism, and 10.0\% (n=4) framework. Forty theories were observed and cited a total of 99 times with the following breakdown: 69.7\% (n=69) general theory, 3.0\% (n=3) principle, 8.1\% (n=8) model, 15.2\% (n=15) mechanism and 4.0\% (n=4) framework. While we found a diversity of theories, fewer than half (46.8\%, n=51) of the papers mentioned even one theory. In those that contained at least one theory, 54.9\% (n=28) referred to two theories, 27.5\% (n=14) mentioned three theories, and 7.8\% (n=4) gave four. 3.9\% (n=2) and 2.0\% (n=1) mentioned five or six theories. Of those mentioning theory, 53\% (n=53) referenced theories applied in the research or described in detail, i.e., \emph{core} theories, while 46.5\% (n=46) did not apply theories or merely mentioned theories without detailed explanation, i.e., \emph{cursory} theories. Sixteen papers only mentioned name of the theory without any further information. One stated use of theory but did not directly reference and describe this usage. One record used a taxonomy as a theory.

\subsubsection{General Theories} More than half (54.9\%, n=28) of papers explicitly referenced Nudge theory. Five used Thaler \& Sunstein's~\cite[p. 8]{ThalerSunstein08} definition: ``any aspect of the choice architecture that alters people's behavior in a predictable way without forbidding any options or significantly changing their economic incentives.'' 21.6\% (n=11) mentioned Dual Process~\cite{kahneman2011thinking}. 11.8\% (n=6) mentioned Choice Architect~\cite{ThalerSunstein08}. 
The restant and our systematic framework of theory name, definition, and the examples of how they were referenced, are in \autoref{tab:theoriesp1} and \autoref{tab:theoriesp2}. This framework shows the diversity of theories, and which theories can be most commonly found, in DP research, and how those theories were used.

\begin{table*}[ht!]
\caption{Theory Used in Dark Patterns Research (Part 1)}
\label{tab:theoriesp1}
\begin{tabular}{p{0.23\linewidth}p{0.35\linewidth}p{0.35\linewidth}}
\toprule
Theory Name (Count)&Definition &Example\\
\midrule
 Nudge (28)& Changes behavior in a predictable way without forbidding any options or significantly changing economic incentives.&``Established on former behavioral economists' ideas ... the idea of `nudge' ...'' ~\cite[p. 4]{nudgeexample}\\
 Dual Process: Automatic \& Reflective System (11)& Distinguishing between two thinking systems: fast/intuitive and slow/analytical.&``One way to explain it is through
the dual-process theories ...''~\cite[p. 4698]{Kitkowska2022}\\
 Choice Architect (6)& The responsibility for organizing the context in which people make decisions&``Thaler and Sunstein ... how intentionally designed choice architectures can nudge people ...''~\cite[p. 377:3]{choicearchitecturesexample}\\
  Sludge (3)& Any aspect of choice architecture where friction makes it hard for people to obtain the best outcome for them &``Richard Thaler refers to dark patterns as `sludge' ...'' ~\cite[p. 377:3]{sludgeexample}\\
 Prospect (3)& A descriptive model of
decision making under risk for creating alternative models&``...  which states human beings' risk-taking willingness ... '' ~\cite[p. 2]{nudgeexample}\\
     Self-Determination Theory (2)&  Competence, relatedness, and autonomy are vital for optimal functioning, growth, and positive social development.&``According to self-determination theory, autonomy is one of the three ...'' ~\cite[p. 52]{selfdeterminationexample}\\
     Proxemics (2)&  Individuals use of space as a specialized elaboration of culture Contex e.g.intimate, personal, social, and public zones.&``The particular innovation we are concerned with is proxemic
interactions ...''~\cite[p. 523]{proxrmicexample}\\
  Privacy Calculus (2)&  A cost-benefit analysis/calculus when deciding whether to disclose private information in various contexts. &``... privacy calculus theory, which presumes people's behaviour ... ''\cite[p. 2]{privacycexample}\\
 Persuasion (2)& Influence is often exerted through six principles: consistency, reciprocation, social proof, authority, liking, and scarcity.&``...social proof, scarcity, authority, ... reciprocation were used ...'' ~\cite[p. 4]{wu2022malicious}\\
  Theory of Planned Behaviour (1)&  Individuals engage in a behavior if they perceive that the benefits of success outweigh the drawbacks of failure.&``According to the theory of planned behavior ...''  ~\cite[p. 5]{KIM2023104763}\\
  Social Influence (1) & Examines how individuals are influenced by others; proposes compliance, identification, and internalization as three main types of social influence. &``...  states that individuals develop their own opinion''~\cite[p. 4]{KIM2023104763}\\
  \bottomrule
  &  &\\
\end{tabular}
\end{table*}

\begin{table*}[ht!]
\caption{Theory Used in Dark Patterns Research (Part 2)}
\label{tab:theoriesp2}
\begin{tabular}{p{0.23\linewidth}p{0.35\linewidth}p{0.35\linewidth}}
\toprule
Theory Name (Count)&Definition &Example\\
\midrule
  Social Control (1)& Delinquency arises when an person's bond to society is weak; includes attachment, commitment, involvement, belief&``...  the social control theory
focused on the community'' ~\cite[p. 3]{socialcontrolexample}\\
  Psychological Reactance (1)&  Individuals are motivated to regain or protect their behavior freedoms when these freedoms are reduced or threatened.&``... the psychological reactance theory suggests ...''~\cite[p. 52]{selfdeterminationexample}\\
 Protection-Motivation (1)& A fear appeal's effectiveness relies on the severity of a depicted event, the likelihood of its occurrence, and the efficacy of a protective response. &``Guided by the Protection-Motivation Theory ...''~\cite[p. 2]{lu2023awareness}\\
  Ludeme (1) &  A minimal element in game design consisting of a grapheme, an acousteme, and a motifeme.&``...  a proposition of a digital game grammar with the ludeme as the `basic video game unit' ...'' ~\cite[p. 10-11]{ludemy}\\
 Information Fiduciary (1)& People and organizations who, because of their relationship with another, assumes special duties with respect to the information they obtain in the course of the relationship. &``... Balkin and Zittrain's information fiduciary theory ...''~\cite[p. 95]{informationfiduciariesexample}\\
  Grounded (1) &  A general methodology for developing theory that is grounded in data systematically gathered and analyzed&``... conducted an open coding process derived from the grounded theory method ...''~\cite[p. 11]{wu2022malicious}\\
  Cognitive Development (1)& Key processes in children's cognition development through their play, dreams, and imitation.&``According to  Piaget's theory of cognitive development ...''~\cite[p. 599]{sousa2023}\\
  \bottomrule
  &  &\\
\end{tabular}
\end{table*}

\subsubsection{Principles}
These were less frequently referenced in the literature, with only three principles mentioned in DP research. 
1.96\% (n=1) each referenced one of Usability Heuristics~\cite{UsabilityHeuristics}, Social proof~\cite{cialdini2009influence,lun2007} and Dual Entitlement~\cite{DualEntitlement}. Details can be found in Supplementary Materials Appendix B.

\subsubsection{Models}
3.92\% (n=2) records referenced Fogg's Behavioral
Model~\cite{foggbehavior} and Mental model~\cite{mentalmodel}. 1.96\% (n=1) each referenced Persuasion knowledge model~\cite{persuasionknowledge}, Affect-Behavior-Cognition model~\cite{Ellis_1991}, Big-five model~\cite{bigfive,bigfive2} or Self-control model~\cite{self-control}. Details can be found in Supplementary Materials Appendix C.

\subsubsection{Frameworks}
These are also infrequently referenced in the literature; only four frameworks were mentioned. 1.96\% (n=1) each referenced Stimulus-Organism-Response framework~\cite{sor,sor2}, Semiotic framework~\cite{Semiotic}, Capability, Opportunity, Motivation-Behavior framework~\cite{COMB} or Behavior Change Wheel framework~\cite{COMB}. Details can be found in Supplementary Materials Appendix D.

\subsubsection{Mechanisms}
These were slightly more frequent. 13.7\% (n=7) referenced the Strategies of Persuasion mechanism~\cite{fogg2003persuasive}. 3.92\% (n=2) referenced the Hyperbolic Discounting mechanism~\cite{HyperbolicDiscounting}. 
Details can be found in Supplementary Materials Appendix E.

\section{Discussion}
Within the community of DP research, there is no consensus on what qualifies as or lies beyond the scope of theory. We have leaned on a broad conceptualization of theory here, to ensure a comprehensive review. Our main finding is that the community could benefit from establishing a uniform definition of what constitutes a theory for DP research. In short, DP scholarship should offer specific theory/ies when referring to ``theory.'' This practice would enhance the application and understanding of theory in the context of DP studies.
As our initial contribution to this process, we first discuss three launching points for future research based on the state of affairs indicated by our scoping review, and then suggest ways forward to deepen theorizing around DPs.

\subsection{Uncovering the State of Affairs: Three Launching Points}

\subsubsection{Point 1: Overall Undertheorizing}
Theory represents a foundation in social science research by providing a systematic and structured approach to understanding, explaining, and predicting social phenomena. ~\citet[p. 173]{abend2008theory} wrote that ``theory is one of the most important words in the lexicon of contemporary sociology,'' going on to explain that the ``words `theory,' `theoretical,' and `theorize' are constantly and consequentially used by all sociologists.'' Yet, DP research is under-theorized and potentially limited in scope given that less than half of the reviewed papers (46.8\%, n=51) reported on an explicit theory. Recent work is more focused on practical research rather than offering theories about phenomena.

Our review also reveals that, apart from Nudge theory~\cite{ThalerSunstein08}, there is a noticeable absence of major theories. One-third of the papers cited Nudge theory (n=28)~\cite{ThalerSunstein08}, advocating for altering people's behavior in a predictable way without forbidding any options or significantly changing their economic incentives. 
Dual Process theory~\cite{kahneman2011thinking} (n=11) was a far second, and posits that individuals think and act under dual automatic and reflective systems. 
Some research referenced Choice Architect theory (n=6), which emphasizes the influence and responsibility of the individual(s) organizing the context in which people make decisions~\cite{ThalerSunstein08}. Seven studies referenced Strategies of Persuasion mechanism which elaborates how seven strategies are commonly used in persuasive technology~\cite{fogg2003persuasive}. A few utilized Sludge theory (n=3), a concept linked to the misuse of nudge~\cite{sludge1}, or Prospect theory (n=3), which explains that people dislike losses more than they like gains~\cite{prospecttheory}. These theories are associated with behavioral economics and psychology, highlighting that DP research is significantly influenced by the broader body of research in human behavior and psychology. The remaining theories were referenced less than three times. 
This raises opportunities to explore other theories within and beyond these larger disciplines.\par

\subsubsection{Point 2: Insufficient Detail and Indirect Citations}
We observed that 46.5\% of papers which referenced theories (n=46) did not explicitly cite or appear to apply these theories. In some instances (n=16), terms such as `theory' or `principle' were mentioned without detail. For example: ``In line with theory on the cute aesthetic, the vast majority of the new wave of home robots appear stunted in some way''~\cite[p. 377]{Lacey2019}. Similarly, some work only contained references to other work that referenced/used the theory in question, i.e., indirect citations. An example: ``The model of van Wynsberghe et al. provides one such model of ethics in design practice, creating a space for ethical theory to be taken up in a pragmatic and generative, rather than objective and static way''~\cite[p. 534]{gray2018} referencing~\cite{vanWynsberghe2014-VANEAD-2}. Short paper formats and work where theory was not the focus may help explain these patterns. Still, they also reveal a broad recognition that theory is important even if overlooked. Indeed, there is a pressing need to enhance the incorporation of theory into future DP research. This would offer a systematic and structured approach to comprehending, explaining, and predicting the social phenomena of DPs, shifting the emphasis away from solely practical research.

\subsubsection{Point 3: Accidental Untheorizing}
What is ``theory''? This is a trickier question than it appears. Authors often use the same terms to describe different epistemological phenomena. ``Framework'' is perhaps infamous. We found two papers in which ``theory'' may have been misapplied. One made reference but did not directly cite or appear to use any theory: ``The aim of this research is to cover this gap of knowledge by using the theory about dark patterns ...''~\cite[p.3]{Bergesen2021}. The other referred to a taxonomy as a theory: ``The analytical framework encompassed two main components: the application of an adapted approach of dark patterns heuristics and theory ...''~\cite[p. 600]{sousa2023}, citing \cite{Dahlan2022FindingDP,Zagal2013DarkPI}. To be fair, it is unclear whether the referenced taxonomy should be classified as theory. 

\subsection{Theorizing a Way Forward: General, Broad, and Specific}

The limited theorization in the existing DP research literature relies, such as it is, very heavily on the psychological theories of Nudge~\cite{ThalerSunstein08}, and to a more limited extent on Fast and Slow Thinking~\cite{kahneman2011thinking}. While psychology should play a key part in providing a theoretical foundation for DP research, it is not the only suitable source. In addition, these two theories focus solely on the psychology of the user, completely ignoring the other side: the psychology of the designers. Here, we highlight some broad theoretical backgrounds which seem to hold promise for providing a broad and solid base for contextualising and informing future DP research. We present these at three levels. First, there is the question of the \emph{general} theoretical framework. This leads to a set of \emph{broad} theories and related bodies of work, from which some \emph{specific} detailed theories can be drawn and specialised for DP research.

\subsubsection{General}

DP research is dependent on a \emph{theory of mind} to explain why users act in ways that are not in their own interests and examine whether that action is deliberately sought by the designer of the system (or not). This requires theorizing about the mental models of users and designers. Theories of mind span the literature in philosophy and (social, behavioral, and neuro-) psychology. The key element for DP research may be \emph{autonomy}: how autonomous we should expect or seek people to be, and the consequences, both individually and societally, have been a subject of philosophical enquiry for millennia. It is one of the key elements of the political theories of Hobbes~\cite{hobbes1651}, Locke~\cite{locke1689second}, and Rawls~\cite{rawls2005}. While philosophy gives a frame for the regulation, or its lack, of DPs in use, it is to psychological theories of autonomy that DP researchers need to turn to understand how DPs work. Much social psychology addresses theories of how people's autonomy is undermined, and how their a-priori values and self-expectations may fail to be followed in real world situations. As Adams~\cite{adams2014facebook} explained in considering the influence of social network sites' core policies around privacy, standard social psychological theories~\cite{Asch1951,bond1996culture,bargh1996automaticity} provide solid frames in which to understand the self-harming choices made by users, and empirical approaches to confirming the reasons for those choices.

\subsubsection{Broad}

This theoretical frame of autonomy from philosophy and psychology tends to focus primarily on the mind of the user. Yet, DP research also needs to draw work on theories of sociology, and in particular of \emph{criminology} and \emph{victimology}, which focus on the social and psychological contexts of perpetrators and victims of crime and other anti-social activity. While most DP usage so far does not rise to the level of crime itself, DPs are used in the furtherance of anti-social online behavior and in support of fraudulent, even criminal, activity. These philosophical frames could be usefully deployed to model and assess level of vicitmization and level of criminology, which may be entwined or not, and may vary across individuals as well as specific types or instances of DPs.

Investigating whether patterns leading users into unwanted activity are \emph{deliberate or inadvertent on behalf of designers} will be a key element in regulating DP deployment. If the designer (humans and organisations) has an ``innocent mind''~\cite{kleinfeld2021mind}, then identifying the deceptive impact of their designs and providing them with a way of avoiding it should be sufficient. If the designer is deliberately deceiving the user, however, public pressure, at the very least, or perhaps regulation will be necessary to remove or ameliorate the deceptive design. For corporate actors (who are perhaps the majority of deployers of DPs), understanding their contextual frame is useful, along the lines of Lessig's~\emph{Code} \cite{lessig2009code} (Laws, Norms, Market, and Architecture as regulators) or Paternoster and Simpson's \emph{Rational Choice}~\cite{paternoster2017rational} (rational choices that consider all information and are judged based on risk vs. reward, favouring reward over risk).

\subsubsection{Specific}

Finally, there are specific theoretical frames in which empirical studies can be designed and justified. Some have been taken up already, as this work shows; there are otherwise too many to list here. The choice should be geared around the motivation of the work. Still, we did find opportunities for greater specificity. For example, some papers reviewed in this study referenced Cialdini~\cite{cialdini2009influence}. A more specific theoretical framework for understanding the victims of fraud developed by Stajano and Wilson~\cite{stajano2011understanding}, which builds on the theory of Cialdini, offers a concrete structure through which to consider how users fall prey to deceptive design. This kind of theory can aid researchers in showing how users can protect themselves, how helpful designers can aid them (by avoiding luring users or by providing them with additional tools to identify and avoid the traps), and how regulators can enforce better practice.


\subsection{Limitations}
We restricted our scoping review to records that used ``dark patterns'' and ``theory'' as keywords in the title, abstract, and/or full-text. We recognize that not all viable papers may have done so. We also note that we may not have found theories that were implied but not explicitly stated in the text. The term ``dark patterns'' is also shifting to other terms, notably ``deceptive designs,'' so further analysis of the existing literature will need to include
this shift in terminology. Since we only included the papers written in English, the results may have demographic biases. We also did not perform a quality assessment as recommended by the PRISMA-Sc protocol because we were not evaluating the results of the included studies.

\section{Conclusion}
Dark patterns and deceptive user interface designs are a widespread phenomenon 
and a sensitive topic from the point of view of experts and purveyors. Here, we have shown the rather limited ways in which this mode of praxis has been approached within the HCI literature from a theoretical lens. We aim to raise awareness and angle the light onto matters of theory in future research. We argue that theory may guide designers, developers, researchers, industry professionals, lawyers, regulatory bodies, and virtually all stakeholders towards consensus and deeper understanding of whether and how ``deceptive'' and harmful these user interface patterns may be for the average digital citizen.

\begin{acks}
We thank Gray et al.~\cite{gray2023dpsysreview} for providing an open data set. We also note that Katie Seaborn conscientiously dissents to in-person participation at CHI; read their positionality statement here: \url{https://bit.ly/chi24statement}
\end{acks}

\balance

\bibliographystyle{ACM-Reference-Format}
\bibliography{bib}


\begin{thebibliography}{68}


\ifx \showCODEN    \undefined \def \showCODEN     #1{\unskip}     \fi
\ifx \showDOI      \undefined \def \showDOI       #1{#1}\fi
\ifx \showISBNx    \undefined \def \showISBNx     #1{\unskip}     \fi
\ifx \showISBNxiii \undefined \def \showISBNxiii  #1{\unskip}     \fi
\ifx \showISSN     \undefined \def \showISSN      #1{\unskip}     \fi
\ifx \showLCCN     \undefined \def \showLCCN      #1{\unskip}     \fi
\ifx \shownote     \undefined \def \shownote      #1{#1}          \fi
\ifx \showarticletitle \undefined \def \showarticletitle #1{#1}   \fi
\ifx \showURL      \undefined \def \showURL       {\relax}        \fi
\providecommand\bibfield[2]{#2}
\providecommand\bibinfo[2]{#2}
\providecommand\natexlab[1]{#1}
\providecommand\showeprint[2][]{arXiv:#2}

\bibitem[Abend(2008)]%
        {abend2008theory}
\bibfield{author}{\bibinfo{person}{Gabriel Abend}.} \bibinfo{year}{2008}\natexlab{}.
\newblock \showarticletitle{The Meaning of ‘Theory’}.
\newblock \bibinfo{journal}{\emph{Sociological Theory}} \bibinfo{volume}{26}, \bibinfo{number}{2} (\bibinfo{year}{2008}), \bibinfo{pages}{173--199}.
\newblock
\urldef\tempurl%
\url{https://doi.org/10.1111/j.1467-9558.2008.00324.x}
\showDOI{\tempurl}


\bibitem[Adams(2014)]%
        {adams2014facebook}
\bibfield{author}{\bibinfo{person}{Andrew~A Adams}.} \bibinfo{year}{2014}\natexlab{}.
\newblock \showarticletitle{Facebook code: Social network sites platform affordances and privacy}.
\newblock \bibinfo{journal}{\emph{Journal of Law, Information and Science}} \bibinfo{volume}{23}, \bibinfo{number}{1} (\bibinfo{year}{2014}), \bibinfo{pages}{158--168}.
\newblock


\bibitem[Ahuja and Kumar(2022)]%
        {selfdeterminationexample}
\bibfield{author}{\bibinfo{person}{Sanju Ahuja} {and} \bibinfo{person}{Jyoti Kumar}.} \bibinfo{year}{2022}\natexlab{}.
\newblock \showarticletitle{Conceptualizations of user autonomy within the normative evaluation of dark patterns}.
\newblock \bibinfo{journal}{\emph{Ethics and Information Technology}}  \bibinfo{volume}{24} (\bibinfo{date}{12} \bibinfo{year}{2022}).
\newblock
\urldef\tempurl%
\url{https://doi.org/10.1007/s10676-022-09672-9}
\showDOI{\tempurl}


\bibitem[Asch(1951)]%
        {Asch1951}
\bibfield{author}{\bibinfo{person}{Solomon~E. Asch}.} \bibinfo{year}{1951}\natexlab{}.
\newblock \showarticletitle{Effects of group pressure upon the modification and distortion of judgments}.
\newblock In \bibinfo{booktitle}{\emph{Groups, Leadership, and Men}}, \bibfield{editor}{\bibinfo{person}{H.~Guetzkow}} (Ed.). \bibinfo{publisher}{Carnegie Press}, \bibinfo{address}{New York, NY, US}.
\newblock


\bibitem[Bargh et~al\mbox{.}(1996)]%
        {bargh1996automaticity}
\bibfield{author}{\bibinfo{person}{John~A Bargh}, \bibinfo{person}{Mark Chen}, {and} \bibinfo{person}{Lara Burrows}.} \bibinfo{year}{1996}\natexlab{}.
\newblock \showarticletitle{Automaticity of social behavior: Direct effects of trait construct and stereotype activation on action.}
\newblock \bibinfo{journal}{\emph{Journal of Personality and Social Psychology}} \bibinfo{volume}{71}, \bibinfo{number}{2} (\bibinfo{year}{1996}), \bibinfo{pages}{230}.
\newblock


\bibitem[Bergesen et~al\mbox{.}(2021)]%
        {Bergesen2021}
\bibfield{author}{\bibinfo{person}{Anders Bergesen}, \bibinfo{person}{John Gullaksen}, \bibinfo{person}{Mikael Hanssen}, \bibinfo{person}{Aleksander Karlsson}, {and} \bibinfo{person}{Birgitte Swensson}.} \bibinfo{year}{2021}\natexlab{}.
\newblock \bibinfo{title}{Dark patterns in cookie consent notices --- Norway's 50 most visited websites}.
\newblock
\newblock


\bibitem[Bond and Smith(1996)]%
        {bond1996culture}
\bibfield{author}{\bibinfo{person}{Rod Bond} {and} \bibinfo{person}{Peter~B Smith}.} \bibinfo{year}{1996}\natexlab{}.
\newblock \showarticletitle{Culture and conformity: A meta-analysis of studies using Asch's (1952b, 1956) line judgment task.}
\newblock \bibinfo{journal}{\emph{Psychological Bulletin}} \bibinfo{volume}{119}, \bibinfo{number}{1} (\bibinfo{year}{1996}), \bibinfo{pages}{111}.
\newblock


\bibitem[Bongard-Blanchy et~al\mbox{.}(2021)]%
        {bongard-blanchy}
\bibfield{author}{\bibinfo{person}{Kerstin Bongard-Blanchy}, \bibinfo{person}{Arianna Rossi}, \bibinfo{person}{Salvador Rivas}, \bibinfo{person}{Sophie Doublet}, \bibinfo{person}{Vincent Koenig}, {and} \bibinfo{person}{Gabriele Lenzini}.} \bibinfo{year}{2021}\natexlab{}.
\newblock \showarticletitle{``I Am Definitely Manipulated, Even When I Am Aware of It. It’s Ridiculous!'' - Dark Patterns from the End-User Perspective}. In \bibinfo{booktitle}{\emph{Proceedings of the 2021 ACM Designing Interactive Systems Conference}} (Virtual Event, USA) \emph{(\bibinfo{series}{DIS '21})}. \bibinfo{publisher}{Association for Computing Machinery}, \bibinfo{address}{New York, NY, USA}, \bibinfo{pages}{763–776}.
\newblock
\showISBNx{9781450384766}
\urldef\tempurl%
\url{https://doi.org/10.1145/3461778.3462086}
\showDOI{\tempurl}


\bibitem[Brignull(2019)]%
        {Brignull_2019}
\bibfield{author}{\bibinfo{person}{Harry Brignull}.} \bibinfo{year}{2019}\natexlab{}.
\newblock \bibinfo{title}{Deceptive patterns - home}.
\newblock
\newblock
\urldef\tempurl%
\url{https://www.deceptive.design/}
\showURL{%
\tempurl}


\bibitem[Bösch et~al\mbox{.}(2016)]%
        {Bösch2016}
\bibfield{author}{\bibinfo{person}{Christoph Bösch}, \bibinfo{person}{Benjamin Erb}, \bibinfo{person}{Frank Kargl}, \bibinfo{person}{Henning Kopp}, {and} \bibinfo{person}{Stefan Pfattheicher}.} \bibinfo{year}{2016}\natexlab{}.
\newblock \showarticletitle{Tales from the Dark Side: Privacy Dark Strategies and Privacy Dark Patterns}.
\newblock \bibinfo{journal}{\emph{Proceedings on Privacy Enhancing Technologies}}  \bibinfo{volume}{2016} (\bibinfo{date}{07} \bibinfo{year}{2016}), \bibinfo{pages}{237–254}.
\newblock
\urldef\tempurl%
\url{https://doi.org/10.1515/popets-2016-0038}
\showDOI{\tempurl}


\bibitem[Chordia et~al\mbox{.}(2023)]%
        {socialcontrolexample}
\bibfield{author}{\bibinfo{person}{Ishita Chordia}, \bibinfo{person}{Lena-Phuong Tran}, \bibinfo{person}{Tala~June Tayebi}, \bibinfo{person}{Emily Parrish}, \bibinfo{person}{Sheena Erete}, \bibinfo{person}{Jason Yip}, {and} \bibinfo{person}{Alexis Hiniker}.} \bibinfo{year}{2023}\natexlab{}.
\newblock \showarticletitle{Deceptive Design Patterns in Safety Technologies: A Case Study of the Citizen App}. In \bibinfo{booktitle}{\emph{Proceedings of the 2023 CHI Conference on Human Factors in Computing Systems}} (Hamburg,Germany) \emph{(\bibinfo{series}{CHI '23})}. \bibinfo{publisher}{Association for Computing Machinery}, \bibinfo{address}{New York, NY, USA}, Article \bibinfo{articleno}{193}, \bibinfo{numpages}{18}~pages.
\newblock
\showISBNx{9781450394215}
\urldef\tempurl%
\url{https://doi.org/10.1145/3544548.3581258}
\showDOI{\tempurl}


\bibitem[Cialdini(2009)]%
        {cialdini2009influence}
\bibfield{author}{\bibinfo{person}{R.B. Cialdini}.} \bibinfo{year}{2009}\natexlab{}.
\newblock \bibinfo{booktitle}{\emph{Influence: The Psychology of Persuasion}}.
\newblock \bibinfo{publisher}{Harper Collins e-books}, \bibinfo{address}{New York, NY, USA}.
\newblock
\showISBNx{9780061899874}


\bibitem[Dahlan and Susanty(2022)]%
        {Dahlan2022FindingDP}
\bibfield{author}{\bibinfo{person}{Refal~Pradama Dahlan} {and} \bibinfo{person}{Meredita Susanty}.} \bibinfo{year}{2022}\natexlab{}.
\newblock \showarticletitle{Finding Dark Patterns in Casual Mobile Games Using Heuristic Evaluation}.
\newblock \bibinfo{journal}{\emph{PETIR}} \bibinfo{volume}{15}, \bibinfo{number}{2} (\bibinfo{year}{2022}), \bibinfo{pages}{185–195}.
\newblock
Issue 2.
\urldef\tempurl%
\url{https://api.semanticscholar.org/CorpusID:254353531}
\showURL{%
\tempurl}


\bibitem[Di~Geronimo et~al\mbox{.}(2020)]%
        {DiGeroimo}
\bibfield{author}{\bibinfo{person}{Linda Di~Geronimo}, \bibinfo{person}{Larissa Braz}, \bibinfo{person}{Enrico Fregnan}, \bibinfo{person}{Fabio Palomba}, {and} \bibinfo{person}{Alberto Bacchelli}.} \bibinfo{year}{2020}\natexlab{}.
\newblock \showarticletitle{UI Dark Patterns and Where to Find Them: A Study on Mobile Applications and User Perception}. In \bibinfo{booktitle}{\emph{Proceedings of the 2020 CHI Conference on Human Factors in Computing Systems}} (Honolulu, HI, USA) \emph{(\bibinfo{series}{CHI '20})}. \bibinfo{publisher}{Association for Computing Machinery}, \bibinfo{address}{New York, NY, USA}, \bibinfo{pages}{1–14}.
\newblock
\showISBNx{9781450367080}
\urldef\tempurl%
\url{https://doi.org/10.1145/3313831.3376600}
\showDOI{\tempurl}


\bibitem[Digman(2003)]%
        {bigfive2}
\bibfield{author}{\bibinfo{person}{John Digman}.} \bibinfo{year}{2003}\natexlab{}.
\newblock \showarticletitle{Personality Structure: Emergence of the Five-Factor Model}.
\newblock \bibinfo{journal}{\emph{Annual Review of Psychology}}  \bibinfo{volume}{41} (\bibinfo{date}{11} \bibinfo{year}{2003}), \bibinfo{pages}{417--440}.
\newblock
\urldef\tempurl%
\url{https://doi.org/10.1146/annurev.ps.41.020190.002221}
\showDOI{\tempurl}


\bibitem[Dupont and Malliet(2021)]%
        {ludemy}
\bibfield{author}{\bibinfo{person}{Bruno Dupont} {and} \bibinfo{person}{Steven Malliet}.} \bibinfo{year}{2021}\natexlab{}.
\newblock \showarticletitle{Contextualizing Dark Patterns with the Ludeme Theory: A New Path for Digital Game Literacy?}
\newblock \bibinfo{journal}{\emph{Acta Ludogica}}  \bibinfo{volume}{4} (\bibinfo{year}{2021}), \bibinfo{pages}{4--22}.
\newblock
Issue 1.
\urldef\tempurl%
\url{https://api.semanticscholar.org/CorpusID:266378823}
\showURL{%
\tempurl}


\bibitem[Ellis(1991)]%
        {Ellis_1991}
\bibfield{author}{\bibinfo{person}{Albert Ellis}.} \bibinfo{year}{1991}\natexlab{}.
\newblock \showarticletitle{The revised ABC’s of rational-emotive therapy (RET)}.
\newblock \bibinfo{journal}{\emph{Journal of Rational-Emotive and Cognitive-Behavior Therapy}} \bibinfo{volume}{9}, \bibinfo{number}{3} (\bibinfo{year}{1991}), \bibinfo{pages}{139–172}.
\newblock
\showISSN{1573-6563}
\urldef\tempurl%
\url{https://doi.org/10.1007/bf01061227}
\showDOI{\tempurl}


\bibitem[Fogg(2003)]%
        {fogg2003persuasive}
\bibfield{author}{\bibinfo{person}{B.~J. Fogg}.} \bibinfo{year}{2003}\natexlab{}.
\newblock \bibinfo{booktitle}{\emph{Persuasive Technology: Using Computers to Change What We Think and Do}}.
\newblock \bibinfo{publisher}{Morgan Kaufmann Publishers}, \bibinfo{address}{Massachusetts, US}.
\newblock
\showISBNx{9781558606432}
\showLCCN{2002110617}
\urldef\tempurl%
\url{https://books.google.co.jp/books?id=9nZHbxULMwgC}
\showURL{%
\tempurl}


\bibitem[Fogg(2009)]%
        {foggbehavior}
\bibfield{author}{\bibinfo{person}{B.~J. Fogg}.} \bibinfo{year}{2009}\natexlab{}.
\newblock \showarticletitle{A Behavior Model for Persuasive Design}. In \bibinfo{booktitle}{\emph{Proceedings of the 4th International Conference on Persuasive Technology}} (Claremont, California, USA) \emph{(\bibinfo{series}{Persuasive '09})}. \bibinfo{publisher}{Association for Computing Machinery}, \bibinfo{address}{New York, NY, USA}, Article \bibinfo{articleno}{40}, \bibinfo{numpages}{7}~pages.
\newblock
\showISBNx{9781605583761}
\urldef\tempurl%
\url{https://doi.org/10.1145/1541948.1541999}
\showDOI{\tempurl}


\bibitem[Friestad and Wright(1994)]%
        {persuasionknowledge}
\bibfield{author}{\bibinfo{person}{Marian Friestad} {and} \bibinfo{person}{Peter Wright}.} \bibinfo{year}{1994}\natexlab{}.
\newblock \showarticletitle{{The Persuasion Knowledge Model: How People Cope with Persuasion Attempts}}.
\newblock \bibinfo{journal}{\emph{Journal of Consumer Research}} \bibinfo{volume}{21}, \bibinfo{number}{1} (\bibinfo{date}{06} \bibinfo{year}{1994}), \bibinfo{pages}{1--31}.
\newblock
\showISSN{0093-5301}
\urldef\tempurl%
\url{https://doi.org/10.1086/209380}
\showDOI{\tempurl}
\showeprint{https://academic.oup.com/jcr/article-pdf/21/1/1/50276032/jcr\_21\_1\_1.pdf}


\bibitem[Goldberg(1991)]%
        {bigfive}
\bibfield{author}{\bibinfo{person}{Lewis Goldberg}.} \bibinfo{year}{1991}\natexlab{}.
\newblock \showarticletitle{An Alternative ``Description of Personality'': The Big-Five Factor Structure}.
\newblock \bibinfo{journal}{\emph{Journal of personality and social psychology}}  \bibinfo{volume}{59} (\bibinfo{date}{01} \bibinfo{year}{1991}), \bibinfo{pages}{1216--29}.
\newblock
\urldef\tempurl%
\url{https://doi.org/10.1037//0022-3514.59.6.1216}
\showDOI{\tempurl}


\bibitem[Grassl et~al\mbox{.}(2021)]%
        {privacycexample}
\bibfield{author}{\bibinfo{person}{Paul Grassl}, \bibinfo{person}{Hanna Schraffenberger}, \bibinfo{person}{Frederik Borgesius}, {and} \bibinfo{person}{Moniek Buijzen}.} \bibinfo{year}{2021}\natexlab{}.
\newblock \showarticletitle{Dark and bright patterns in cookie consent requests}.
\newblock \bibinfo{journal}{\emph{Journal of Digital Social Research}} \bibinfo{volume}{3}, \bibinfo{number}{1} (\bibinfo{year}{2021}), \bibinfo{pages}{1--38}.
\newblock
\urldef\tempurl%
\url{https://doi.org/10.31234/osf.io/gqs5h}
\showDOI{\tempurl}


\bibitem[Gray et~al\mbox{.}(2018)]%
        {gray2018}
\bibfield{author}{\bibinfo{person}{Colin~M. Gray}, \bibinfo{person}{Yubo Kou}, \bibinfo{person}{Bryan Battles}, \bibinfo{person}{Joseph Hoggatt}, {and} \bibinfo{person}{Austin~L. Toombs}.} \bibinfo{year}{2018}\natexlab{}.
\newblock \showarticletitle{The Dark (Patterns) Side of UX Design}. In \bibinfo{booktitle}{\emph{Proceedings of the 2018 CHI Conference on Human Factors in Computing Systems}} (Montreal, QC, Canada) \emph{(\bibinfo{series}{CHI '18})}. \bibinfo{publisher}{Association for Computing Machinery}, \bibinfo{address}{New York, NY, USA}, \bibinfo{pages}{1–14}.
\newblock
\showISBNx{9781450356206}
\urldef\tempurl%
\url{https://doi.org/10.1145/3173574.3174108}
\showDOI{\tempurl}


\bibitem[Gray et~al\mbox{.}(2023)]%
        {gray2023dpsysreview}
\bibfield{author}{\bibinfo{person}{Colin~M. Gray}, \bibinfo{person}{Lorena Sanchez~Chamorro}, \bibinfo{person}{Ike Obi}, {and} \bibinfo{person}{Ja-Nae Duane}.} \bibinfo{year}{2023}\natexlab{}.
\newblock \showarticletitle{Mapping the Landscape of Dark Patterns Scholarship: A Systematic Literature Review}. In \bibinfo{booktitle}{\emph{Companion Publication of the 2023 ACM Designing Interactive Systems Conference}} (Pittsburgh, PA, USA) \emph{(\bibinfo{series}{DIS '23 Companion})}. \bibinfo{publisher}{Association for Computing Machinery}, \bibinfo{address}{New York, NY, USA}, \bibinfo{pages}{188–193}.
\newblock
\showISBNx{9781450398985}
\urldef\tempurl%
\url{https://doi.org/10.1145/3563703.3596635}
\showDOI{\tempurl}


\bibitem[Greenberg et~al\mbox{.}(2014)]%
        {proxrmicexample}
\bibfield{author}{\bibinfo{person}{Saul Greenberg}, \bibinfo{person}{Sebastian Boring}, \bibinfo{person}{Jo Vermeulen}, {and} \bibinfo{person}{Jakub Dostal}.} \bibinfo{year}{2014}\natexlab{}.
\newblock \showarticletitle{Dark Patterns in Proxemic Interactions: A Critical Perspective}. In \bibinfo{booktitle}{\emph{Proceedings of the 2014 Conference on Designing Interactive Systems}} (Vancouver, BC, Canada) \emph{(\bibinfo{series}{DIS '14})}. \bibinfo{publisher}{Association for Computing Machinery}, \bibinfo{address}{New York, NY, USA}, \bibinfo{pages}{523–532}.
\newblock
\showISBNx{9781450329026}
\urldef\tempurl%
\url{https://doi.org/10.1145/2598510.2598541}
\showDOI{\tempurl}


\bibitem[Gregor(2006)]%
        {Gregor2006}
\bibfield{author}{\bibinfo{person}{Shirley Gregor}.} \bibinfo{year}{2006}\natexlab{}.
\newblock \showarticletitle{The Nature of Theory in Information Systems}.
\newblock \bibinfo{journal}{\emph{MIS Quarterly}}  \bibinfo{volume}{30} (\bibinfo{date}{09} \bibinfo{year}{2006}), \bibinfo{pages}{611--642}.
\newblock
\urldef\tempurl%
\url{https://doi.org/10.2307/25148742}
\showDOI{\tempurl}


\bibitem[Gunawan et~al\mbox{.}(2021a)]%
        {choicearchitecturesexample}
\bibfield{author}{\bibinfo{person}{Johanna Gunawan}, \bibinfo{person}{Amogh Pradeep}, \bibinfo{person}{David Choffnes}, \bibinfo{person}{Woodrow Hartzog}, {and} \bibinfo{person}{Christo Wilson}.} \bibinfo{year}{2021}\natexlab{a}.
\newblock \showarticletitle{A Comparative Study of Dark Patterns Across Web and Mobile Modalities}.
\newblock \bibinfo{journal}{\emph{Proc. ACM Hum.-Comput. Interact.}} \bibinfo{volume}{5}, \bibinfo{number}{CSCW2}, Article \bibinfo{articleno}{377} (\bibinfo{date}{oct} \bibinfo{year}{2021}), \bibinfo{numpages}{29}~pages.
\newblock
\urldef\tempurl%
\url{https://doi.org/10.1145/3479521}
\showDOI{\tempurl}


\bibitem[Gunawan et~al\mbox{.}(2021b)]%
        {sludgeexample}
\bibfield{author}{\bibinfo{person}{Johanna Gunawan}, \bibinfo{person}{Amogh Pradeep}, \bibinfo{person}{David Choffnes}, \bibinfo{person}{Woodrow Hartzog}, {and} \bibinfo{person}{Christo Wilson}.} \bibinfo{year}{2021}\natexlab{b}.
\newblock \showarticletitle{A Comparative Study of Dark Patterns Across Web and Mobile Modalities}.
\newblock \bibinfo{journal}{\emph{Proc. ACM Hum.-Comput. Interact.}} \bibinfo{volume}{5}, \bibinfo{number}{CSCW2}, Article \bibinfo{articleno}{377} (\bibinfo{date}{oct} \bibinfo{year}{2021}), \bibinfo{numpages}{29}~pages.
\newblock
\urldef\tempurl%
\url{https://doi.org/10.1145/3479521}
\showDOI{\tempurl}


\bibitem[Harry(2023)]%
        {Brignull_2023}
\bibfield{author}{\bibinfo{person}{Brignull Harry}.} \bibinfo{year}{2023}\natexlab{}.
\newblock \bibinfo{booktitle}{\emph{Deceptive Patterns: Exposing the Tricks Tech Companies Use to Control You.}}
\newblock \bibinfo{publisher}{Testimonium Ltd}, \bibinfo{address}{Eastbourne, England}.
\newblock


\bibitem[Hobbes(1651)]%
        {hobbes1651}
\bibfield{author}{\bibinfo{person}{Thomas Hobbes}.} \bibinfo{year}{1651}\natexlab{}.
\newblock \bibinfo{booktitle}{\emph{Hobbes's Leviathan}}.
\newblock \bibinfo{publisher}{Clarendon Press}, \bibinfo{address}{Oxford, United Kingdom}.
\newblock


\bibitem[Hsieh and Shannon(2005)]%
        {contentreview}
\bibfield{author}{\bibinfo{person}{Hsiu-Fang Hsieh} {and} \bibinfo{person}{Sarah Shannon}.} \bibinfo{year}{2005}\natexlab{}.
\newblock \showarticletitle{Three Approaches to Qualitative Content Analysis}.
\newblock \bibinfo{journal}{\emph{Qualitative health research}}  \bibinfo{volume}{15} (\bibinfo{date}{12} \bibinfo{year}{2005}), \bibinfo{pages}{1277--88}.
\newblock
\urldef\tempurl%
\url{https://doi.org/10.1177/1049732305276687}
\showDOI{\tempurl}


\bibitem[Huang(2023)]%
        {sor}
\bibfield{author}{\bibinfo{person}{Tianyang Huang}.} \bibinfo{year}{2023}\natexlab{}.
\newblock \showarticletitle{Using SOR framework to explore the driving factors of older adults smartphone use behavior}.
\newblock \bibinfo{journal}{\emph{Humanities and Social Sciences Communications}}  \bibinfo{volume}{10} (\bibinfo{date}{10} \bibinfo{year}{2023}).
\newblock
\urldef\tempurl%
\url{https://doi.org/10.1057/s41599-023-02221-9}
\showDOI{\tempurl}


\bibitem[Kahneman(2011)]%
        {kahneman2011thinking}
\bibfield{author}{\bibinfo{person}{Daniel Kahneman}.} \bibinfo{year}{2011}\natexlab{}.
\newblock \bibinfo{booktitle}{\emph{Thinking, fast and slow}}.
\newblock \bibinfo{publisher}{Farrar, Straus and Giroux}, \bibinfo{address}{New York, NY, USA}.
\newblock
\showISBNx{9780374275631 0374275637}


\bibitem[Kahneman et~al\mbox{.}(1986)]%
        {DualEntitlement}
\bibfield{author}{\bibinfo{person}{Daniel Kahneman}, \bibinfo{person}{Jack~L. Knetsch}, {and} \bibinfo{person}{Richard~H. Thaler}.} \bibinfo{year}{1986}\natexlab{}.
\newblock \showarticletitle{Fairness and The Assumptions of Economics}.
\newblock \bibinfo{journal}{\emph{The Journal of Business}}  \bibinfo{volume}{59} (\bibinfo{date}{10} \bibinfo{year}{1986}), \bibinfo{pages}{285--300}.
\newblock
\urldef\tempurl%
\url{https://doi.org/10.1086/296367}
\showDOI{\tempurl}


\bibitem[Kahneman and Tversky(1979)]%
        {prospecttheory}
\bibfield{author}{\bibinfo{person}{Daniel Kahneman} {and} \bibinfo{person}{Amos Tversky}.} \bibinfo{year}{1979}\natexlab{}.
\newblock \showarticletitle{{Prospect Theory: An Analysis of Decision under Risk}}.
\newblock \bibinfo{journal}{\emph{Econometrica}} \bibinfo{volume}{47}, \bibinfo{number}{2} (\bibinfo{date}{March} \bibinfo{year}{1979}), \bibinfo{pages}{263--291}.
\newblock
\urldef\tempurl%
\url{https://ideas.repec.org/a/ecm/emetrp/v47y1979i2p263-91.html}
\showURL{%
\tempurl}


\bibitem[Kim et~al\mbox{.}(2023)]%
        {KIM2023104763}
\bibfield{author}{\bibinfo{person}{Kawon~(Kathy) Kim}, \bibinfo{person}{Woo~Gon Kim}, {and} \bibinfo{person}{Minwoo Lee}.} \bibinfo{year}{2023}\natexlab{}.
\newblock \showarticletitle{Impact of dark patterns on consumers’ perceived fairness and attitude: Moderating effects of types of dark patterns, social proof, and moral identity}.
\newblock \bibinfo{journal}{\emph{Tourism Management}}  \bibinfo{volume}{98} (\bibinfo{year}{2023}), \bibinfo{pages}{104763}.
\newblock
\showISSN{0261-5177}
\urldef\tempurl%
\url{https://doi.org/10.1016/j.tourman.2023.104763}
\showDOI{\tempurl}


\bibitem[Kitkowska et~al\mbox{.}(2022)]%
        {Kitkowska2022}
\bibfield{author}{\bibinfo{person}{Agnieszka Kitkowska}, \bibinfo{person}{Johan Högberg}, {and} \bibinfo{person}{Erik Wästlund}.} \bibinfo{year}{2022}\natexlab{}.
\newblock \showarticletitle{Barriers to a Well-Functioning Digital Market: Exploring Dark Patterns and How to Overcome Them}. In \bibinfo{booktitle}{\emph{55th Hawaii International Conference on System Sciences}}. \bibinfo{publisher}{ScholarSpace}, \bibinfo{address}{Hawaii, US}, \bibinfo{pages}{4697--4706}.
\newblock
\urldef\tempurl%
\url{https://doi.org/10.24251/HICSS.2022.573}
\showDOI{\tempurl}


\bibitem[Kleinfeld(2021)]%
        {kleinfeld2021mind}
\bibfield{author}{\bibinfo{person}{Joshua Kleinfeld}.} \bibinfo{year}{2021}\natexlab{}.
\newblock \showarticletitle{Why the mind matters in criminal law}.
\newblock \bibinfo{journal}{\emph{Ariz. St. LJ}}  \bibinfo{volume}{53} (\bibinfo{year}{2021}), \bibinfo{pages}{539}.
\newblock


\bibitem[Lacey and Caudwell(2019)]%
        {Lacey2019}
\bibfield{author}{\bibinfo{person}{Cherie Lacey} {and} \bibinfo{person}{Catherine Caudwell}.} \bibinfo{year}{2019}\natexlab{}.
\newblock \showarticletitle{Cuteness as a ‘Dark Pattern’ in Home Robots}. In \bibinfo{booktitle}{\emph{2019 14th ACM/IEEE International Conference on Human-Robot Interaction (HRI)}}. \bibinfo{publisher}{IEEE Press}, \bibinfo{address}{Daegu, Republic of Korea}, \bibinfo{pages}{374--381}.
\newblock
\showISSN{2167-2148}
\urldef\tempurl%
\url{https://doi.org/10.1109/HRI.2019.8673274}
\showDOI{\tempurl}


\bibitem[Laibson(1997)]%
        {HyperbolicDiscounting}
\bibfield{author}{\bibinfo{person}{David Laibson}.} \bibinfo{year}{1997}\natexlab{}.
\newblock \showarticletitle{Golden Eggs and Hyperbolic Discounting}.
\newblock \bibinfo{journal}{\emph{The Quarterly Journal of Economics}}  \bibinfo{volume}{112} (\bibinfo{date}{02} \bibinfo{year}{1997}), \bibinfo{pages}{443--77}.
\newblock
\urldef\tempurl%
\url{https://doi.org/10.1162/003355397555253}
\showDOI{\tempurl}


\bibitem[Lessig(2009)]%
        {lessig2009code}
\bibfield{author}{\bibinfo{person}{Lawrence Lessig}.} \bibinfo{year}{2009}\natexlab{}.
\newblock \bibinfo{booktitle}{\emph{Code: and Other Laws of Cyberspace}}.
\newblock \bibinfo{publisher}{Basic Books}, \bibinfo{address}{New York, NY, US}.
\newblock


\bibitem[Locke(1689)]%
        {locke1689second}
\bibfield{author}{\bibinfo{person}{John Locke}.} \bibinfo{year}{1689}\natexlab{}.
\newblock \bibinfo{booktitle}{\emph{Second Treatise of Government}}.
\newblock \bibinfo{publisher}{Watchmaker Publishing}, \bibinfo{address}{Oregon, US}.
\newblock


\bibitem[Lu et~al\mbox{.}(2023)]%
        {lu2023awareness}
\bibfield{author}{\bibinfo{person}{Yuwen Lu}, \bibinfo{person}{Chao Zhang}, \bibinfo{person}{Yuewen Yang}, \bibinfo{person}{Yaxing Yao}, {and} \bibinfo{person}{Toby Jia-Jun Li}.} \bibinfo{year}{2023}\natexlab{}.
\newblock \bibinfo{title}{From Awareness to Action: Exploring End-User Empowerment Interventions for Dark Patterns in UX}.
\newblock
\newblock
\showeprint[arxiv]{2310.17846}~[cs.HC]


\bibitem[Luguri and Strahilevitz(2021)]%
        {informationfiduciariesexample}
\bibfield{author}{\bibinfo{person}{Jamie Luguri} {and} \bibinfo{person}{Lior~Jacob Strahilevitz}.} \bibinfo{year}{2021}\natexlab{}.
\newblock \showarticletitle{{Shining a Light on Dark Patterns}}.
\newblock \bibinfo{journal}{\emph{Journal of Legal Analysis}} \bibinfo{volume}{13}, \bibinfo{number}{1} (\bibinfo{date}{03} \bibinfo{year}{2021}), \bibinfo{pages}{43--109}.
\newblock
\showISSN{2161-7201}
\urldef\tempurl%
\url{https://doi.org/10.1093/jla/laaa006}
\showDOI{\tempurl}
\showeprint{https://academic.oup.com/jla/article-pdf/13/1/43/36669915/laaa006.pdf}


\bibitem[Lun et~al\mbox{.}(2007)]%
        {lun2007}
\bibfield{author}{\bibinfo{person}{Janetta Lun}, \bibinfo{person}{Stacey Sinclair}, \bibinfo{person}{Erin Whitchurch}, {and} \bibinfo{person}{Catherine Glenn}.} \bibinfo{year}{2007}\natexlab{}.
\newblock \showarticletitle{(Why) Do I Think What You Think? Epistemic Social Tuning and Implicit Prejudice}.
\newblock \bibinfo{journal}{\emph{Journal of personality and social psychology}}  \bibinfo{volume}{93} (\bibinfo{date}{12} \bibinfo{year}{2007}), \bibinfo{pages}{957--72}.
\newblock
\urldef\tempurl%
\url{https://doi.org/10.1037/0022-3514.93.6.957}
\showDOI{\tempurl}


\bibitem[Machamer et~al\mbox{.}(2000)]%
        {Machamer2000}
\bibfield{author}{\bibinfo{person}{Peter Machamer}, \bibinfo{person}{Lindley Darden}, {and} \bibinfo{person}{Carl~F. Craver}.} \bibinfo{year}{2000}\natexlab{}.
\newblock \showarticletitle{Thinking about Mechanisms}.
\newblock \bibinfo{journal}{\emph{Philosophy of Science}} \bibinfo{volume}{67}, \bibinfo{number}{1} (\bibinfo{year}{2000}), \bibinfo{pages}{1–25}.
\newblock
\urldef\tempurl%
\url{https://doi.org/10.1086/392759}
\showDOI{\tempurl}


\bibitem[McDonald(2009)]%
        {McDonald2009}
\bibfield{author}{\bibinfo{person}{Hugh~P. McDonald}.} \bibinfo{year}{2009}\natexlab{}.
\newblock \showarticletitle{Principles: The Principles of Principles}.
\newblock \bibinfo{journal}{\emph{The Pluralist}} \bibinfo{volume}{4}, \bibinfo{number}{3} (\bibinfo{year}{2009}), \bibinfo{pages}{98--126}.
\newblock
\urldef\tempurl%
\url{https://doi.org/10.1353/plu.0.0028}
\showDOI{\tempurl}


\bibitem[Mehrabian and Russell(1974)]%
        {sor2}
\bibfield{author}{\bibinfo{person}{A. Mehrabian} {and} \bibinfo{person}{J.A. Russell}.} \bibinfo{year}{1974}\natexlab{}.
\newblock \bibinfo{booktitle}{\emph{An Approach to Environmental Psychology}}.
\newblock \bibinfo{publisher}{M.I.T. Press}, \bibinfo{address}{Massachusetts, US}.
\newblock
\showISBNx{9780262130905}
\showLCCN{73016437}


\bibitem[Metcalfe and Mischel(1999)]%
        {self-control}
\bibfield{author}{\bibinfo{person}{Janet Metcalfe} {and} \bibinfo{person}{Walter Mischel}.} \bibinfo{year}{1999}\natexlab{}.
\newblock \showarticletitle{A Hot/Cool-System Analysis of Delay of Gratification: Dynamics of Willpower}.
\newblock \bibinfo{journal}{\emph{Psychological review}}  \bibinfo{volume}{106} (\bibinfo{date}{01} \bibinfo{year}{1999}), \bibinfo{pages}{3--19}.
\newblock
\urldef\tempurl%
\url{https://doi.org/10.1037/0033-295X.106.1.3}
\showDOI{\tempurl}


\bibitem[Michie et~al\mbox{.}(2011)]%
        {COMB}
\bibfield{author}{\bibinfo{person}{Susan Michie}, \bibinfo{person}{Maartje van Stralen}, {and} \bibinfo{person}{Robert West}.} \bibinfo{year}{2011}\natexlab{}.
\newblock \showarticletitle{The Behaviour Change Wheel: a new method for characterising and designing behaviour change interventions}.
\newblock \bibinfo{journal}{\emph{Implementation science : IS}}  \bibinfo{volume}{6} (\bibinfo{date}{04} \bibinfo{year}{2011}), \bibinfo{pages}{42}.
\newblock
\urldef\tempurl%
\url{https://doi.org/10.1186/1748-5908-6-42}
\showDOI{\tempurl}


\bibitem[Nielsen(1994)]%
        {UsabilityHeuristics}
\bibfield{author}{\bibinfo{person}{Jakob Nielsen}.} \bibinfo{year}{1994}\natexlab{}.
\newblock \showarticletitle{Enhancing the Explanatory Power of Usability Heuristics}. In \bibinfo{booktitle}{\emph{Proceedings of the SIGCHI Conference on Human Factors in Computing Systems}} (Boston, Massachusetts, USA) \emph{(\bibinfo{series}{CHI '94})}. \bibinfo{publisher}{Association for Computing Machinery}, \bibinfo{address}{New York, NY, USA}, \bibinfo{pages}{152–158}.
\newblock
\showISBNx{0897916506}
\urldef\tempurl%
\url{https://doi.org/10.1145/191666.191729}
\showDOI{\tempurl}


\bibitem[Nilsen(2015)]%
        {nilsen2015}
\bibfield{author}{\bibinfo{person}{Per Nilsen}.} \bibinfo{year}{2015}\natexlab{}.
\newblock \showarticletitle{Making sense of implementation theories, models and frameworks}.
\newblock \bibinfo{journal}{\emph{Implementation science : IS}}  \bibinfo{volume}{10} (\bibinfo{date}{04} \bibinfo{year}{2015}), \bibinfo{pages}{53}.
\newblock
\urldef\tempurl%
\url{https://doi.org/10.1186/s13012-015-0242-0}
\showDOI{\tempurl}


\bibitem[Norman(2002)]%
        {mentalmodel}
\bibfield{author}{\bibinfo{person}{Donald~A. Norman}.} \bibinfo{year}{2002}\natexlab{}.
\newblock \bibinfo{booktitle}{\emph{The Design of Everyday Things}}.
\newblock \bibinfo{publisher}{Basic Books, Inc.}, \bibinfo{address}{USA}.
\newblock
\showISBNx{9780465067107}


\bibitem[OECD(2022)]%
        {oecd2022}
\bibfield{author}{\bibinfo{person}{OECD}.} \bibinfo{year}{2022}\natexlab{}.
\newblock \bibinfo{booktitle}{\emph{Dark commercial patterns}}.
\newblock Number 336 in \bibinfo{series}{OECD Digital Economy Papers}. \bibinfo{publisher}{OECD Publishing}, \bibinfo{address}{Paris, France}. 96 pages.
\newblock
\urldef\tempurl%
\url{https://doi.org/10.1787/44f5e846-en}
\showDOI{\tempurl}


\bibitem[Ozdemir(2019)]%
        {nudgeexample}
\bibfield{author}{\bibinfo{person}{Sebnem Ozdemir}.} \bibinfo{year}{2019}\natexlab{}.
\newblock \showarticletitle{Digital nudges and dark patterns: The angels and the archfiends of digital communication}.
\newblock \bibinfo{journal}{\emph{Digital Scholarship in the Humanities}}  \bibinfo{volume}{35} (\bibinfo{date}{03} \bibinfo{year}{2019}).
\newblock
\urldef\tempurl%
\url{https://doi.org/10.1093/llc/fqz014}
\showDOI{\tempurl}


\bibitem[Paternoster and Simpson(2017)]%
        {paternoster2017rational}
\bibfield{author}{\bibinfo{person}{Raymond Paternoster} {and} \bibinfo{person}{Sally Simpson}.} \bibinfo{year}{2017}\natexlab{}.
\newblock \showarticletitle{A rational choice theory of corporate crime}.
\newblock In \bibinfo{booktitle}{\emph{Routine Activity and Rational Choice}}, \bibfield{editor}{\bibinfo{person}{Ronald Victor~Gemuseus Clarke} {and} \bibinfo{person}{Marcus Felson}} (Eds.). \bibinfo{publisher}{Routledge}, \bibinfo{address}{London, UK}, \bibinfo{pages}{37--58}.
\newblock


\bibitem[Rawls(2005)]%
        {rawls2005}
\bibfield{author}{\bibinfo{person}{John Rawls}.} \bibinfo{year}{2005}\natexlab{}.
\newblock \bibinfo{booktitle}{\emph{Political Liberalism}}.
\newblock \bibinfo{publisher}{Columbia University Press}, \bibinfo{address}{New York, NY, US}.
\newblock
\newblock
\shownote{Expanded edition from 1993 original}.


\bibitem[Sousa and Oliveira(2023)]%
        {sousa2023}
\bibfield{author}{\bibinfo{person}{Carla Sousa} {and} \bibinfo{person}{Ana Oliveira}.} \bibinfo{year}{2023}\natexlab{}.
\newblock \showarticletitle{The Dark Side of Fun: Understanding Dark Patterns and Literacy Needs in Early Childhood Mobile Gaming}, In \bibinfo{booktitle}{Vol. 17 No. 1 (2023): Proceedings of the 17th European Conference on Games Based Learning}.
\newblock \bibinfo{journal}{\emph{European Conference on Games Based Learning}}  \bibinfo{volume}{17}.
\newblock
\urldef\tempurl%
\url{https://doi.org/10.34190/ecgbl.17.1.1656}
\showDOI{\tempurl}


\bibitem[Stajano and Wilson(2011)]%
        {stajano2011understanding}
\bibfield{author}{\bibinfo{person}{Frank Stajano} {and} \bibinfo{person}{Paul Wilson}.} \bibinfo{year}{2011}\natexlab{}.
\newblock \showarticletitle{Understanding scam victims: Seven principles for systems security}.
\newblock \bibinfo{journal}{\emph{Commun. ACM}} \bibinfo{volume}{54}, \bibinfo{number}{3} (\bibinfo{year}{2011}), \bibinfo{pages}{70--75}.
\newblock


\bibitem[Stamper(1993)]%
        {Semiotic}
\bibfield{author}{\bibinfo{person}{Ronald~K. Stamper}.} \bibinfo{year}{1993}\natexlab{}.
\newblock \showarticletitle{A semiotic theory of information and information systems}. In \bibinfo{booktitle}{\emph{Proceedings of the Joint ICL/University of Newcastle Seminar on the Teaching of Computer Science, Part IX: Information}}. \bibinfo{publisher}{University of Newcastle}, \bibinfo{address}{Newcastle, UK}, \bibinfo{numpages}{1-33}~pages.
\newblock
\urldef\tempurl%
\url{https://api.semanticscholar.org/CorpusID:64006127}
\showURL{%
\tempurl}


\bibitem[Sutton and Staw(1995)]%
        {notheory}
\bibfield{author}{\bibinfo{person}{Robert~I. Sutton} {and} \bibinfo{person}{Barry~M. Staw}.} \bibinfo{year}{1995}\natexlab{}.
\newblock \showarticletitle{What Theory is Not}.
\newblock \bibinfo{journal}{\emph{Administrative Science Quarterly}} \bibinfo{volume}{40}, \bibinfo{number}{3} (\bibinfo{year}{1995}), \bibinfo{pages}{371--384}.
\newblock
\showISSN{00018392}
\urldef\tempurl%
\url{http://www.jstor.org/stable/2393788}
\showURL{%
\tempurl}


\bibitem[Thaler(2018)]%
        {sludge1}
\bibfield{author}{\bibinfo{person}{Richard~H. Thaler}.} \bibinfo{year}{2018}\natexlab{}.
\newblock \showarticletitle{Nudge, not sludge}.
\newblock \bibinfo{journal}{\emph{Science}} \bibinfo{volume}{361}, \bibinfo{number}{6401} (\bibinfo{year}{2018}), \bibinfo{pages}{431--431}.
\newblock
\urldef\tempurl%
\url{https://doi.org/10.1126/science.aau9241}
\showDOI{\tempurl}
\showeprint{https://www.science.org/doi/pdf/10.1126/science.aau9241}


\bibitem[Thaler and Sunstein(2008)]%
        {ThalerSunstein08}
\bibfield{author}{\bibinfo{person}{Richard~H. Thaler} {and} \bibinfo{person}{Cass~R. Sunstein}.} \bibinfo{year}{2008}\natexlab{}.
\newblock \bibinfo{booktitle}{\emph{Nudge}}.
\newblock \bibinfo{publisher}{Yale University Press}, \bibinfo{address}{New Haven, CT and London}.
\newblock
\showISBNx{978-0-300-12223-7}


\bibitem[Tricco et~al\mbox{.}(2018)]%
        {PRISMA_checklist}
\bibfield{author}{\bibinfo{person}{Andrea~C. Tricco}, \bibinfo{person}{Erin Lillie}, \bibinfo{person}{Wasifa Zarin}, \bibinfo{person}{Kelly~K. O’Brien}, \bibinfo{person}{Heather Colquhoun}, \bibinfo{person}{Danielle Levac}, \bibinfo{person}{David Moher}, \bibinfo{person}{Micah~D.J. Peters}, \bibinfo{person}{Tanya Horsley}, \bibinfo{person}{Laura Weeks}, \bibinfo{person}{Susanne Hempel}, \bibinfo{person}{Elie~A. Akl}, \bibinfo{person}{Christine Chang}, \bibinfo{person}{Jessie McGowan}, \bibinfo{person}{Lesley Stewart}, \bibinfo{person}{Lisa Hartling}, \bibinfo{person}{Adrian Aldcroft}, \bibinfo{person}{Michael~G. Wilson}, \bibinfo{person}{Chantelle Garritty}, \bibinfo{person}{Simon Lewin}, \bibinfo{person}{Christina~M. Godfrey}, \bibinfo{person}{Marilyn~T. Macdonald}, \bibinfo{person}{Etienne~V. Langlois}, \bibinfo{person}{Karla Soares-Weiser}, \bibinfo{person}{Jo Moriarty}, \bibinfo{person}{Tammy Clifford}, \bibinfo{person}{\"{O}zge Tun\c{c}alp}, {and} \bibinfo{person}{Sharon~E. Straus}.}
  \bibinfo{year}{2018}\natexlab{}.
\newblock \showarticletitle{PRISMA Extension for Scoping Reviews (PRISMA-ScR): Checklist and Explanation}.
\newblock \bibinfo{journal}{\emph{Annals of Internal Medicine}} \bibinfo{volume}{169}, \bibinfo{number}{7} (\bibinfo{year}{2018}), \bibinfo{pages}{467--473}.
\newblock
\urldef\tempurl%
\url{https://doi.org/10.7326/M18-0850}
\showDOI{\tempurl}
\showeprint{https://doi.org/10.7326/M18-0850}
\newblock
\shownote{PMID: 30178033}.


\bibitem[van Wynsberghe and Robbins(2014)]%
        {vanWynsberghe2014-VANEAD-2}
\bibfield{author}{\bibinfo{person}{Aimee van Wynsberghe} {and} \bibinfo{person}{Scott Robbins}.} \bibinfo{year}{2014}\natexlab{}.
\newblock \showarticletitle{Ethicist as Designer: A Pragmatic Approach to Ethics in the Lab}.
\newblock \bibinfo{journal}{\emph{Science and Engineering Ethics}} \bibinfo{volume}{20}, \bibinfo{number}{4} (\bibinfo{year}{2014}), \bibinfo{pages}{947--961}.
\newblock
\urldef\tempurl%
\url{https://doi.org/10.1007/s11948-013-9498-4}
\showDOI{\tempurl}


\bibitem[Wacker(1998)]%
        {wacker1998definition}
\bibfield{author}{\bibinfo{person}{John~G Wacker}.} \bibinfo{year}{1998}\natexlab{}.
\newblock \showarticletitle{A definition of theory: research guidelines for different theory-building research methods in operations management}.
\newblock \bibinfo{journal}{\emph{Journal of Operations Management}} \bibinfo{volume}{16}, \bibinfo{number}{4} (\bibinfo{year}{1998}), \bibinfo{pages}{361--385}.
\newblock


\bibitem[Wu et~al\mbox{.}(2023)]%
        {wu2022malicious}
\bibfield{author}{\bibinfo{person}{Qunfang Wu}, \bibinfo{person}{Yisi Sang}, \bibinfo{person}{Dakuo Wang}, {and} \bibinfo{person}{Zhicong Lu}.} \bibinfo{year}{2023}\natexlab{}.
\newblock \showarticletitle{Malicious Selling Strategies in Livestream E-commerce: A Case Study of Alibaba’s Taobao and ByteDance’s TikTok}.
\newblock \bibinfo{journal}{\emph{ACM Trans. Comput.-Hum. Interact.}} \bibinfo{volume}{30}, \bibinfo{number}{3}, Article \bibinfo{articleno}{35} (\bibinfo{date}{jun} \bibinfo{year}{2023}), \bibinfo{numpages}{29}~pages.
\newblock
\showISSN{1073-0516}
\urldef\tempurl%
\url{https://doi.org/10.1145/3577199}
\showDOI{\tempurl}


\bibitem[Zagal et~al\mbox{.}(2013)]%
        {Zagal2013DarkPI}
\bibfield{author}{\bibinfo{person}{Jos{\'e}~Pablo Zagal}, \bibinfo{person}{Staffan Bj{\"o}rk}, {and} \bibinfo{person}{Chris Lewis}.} \bibinfo{year}{2013}\natexlab{}.
\newblock \showarticletitle{Dark patterns in the design of games}. In \bibinfo{booktitle}{\emph{International Conference on Foundations of Digital Games}}. \bibinfo{publisher}{{}}, \bibinfo{address}{Crete, Greece}.
\newblock
\urldef\tempurl%
\url{https://api.semanticscholar.org/CorpusID:17683705}
\showURL{%
\tempurl}


\end{thebibliography}

\end{document}